\documentclass[runningheads]{llncs}

\usepackage[T1]{fontenc}
\usepackage{graphicx}
\usepackage{hyperref}
\usepackage{color}
\usepackage[colorinlistoftodos]{todonotes}
\usepackage{amsmath}
\usepackage{booktabs}

\DeclareUnicodeCharacter{2212}{-}

\begin{document}

\title{Towards Effective Immersive Technologies in Medicine: Potential and Future Applications based on VR, AR, XR and AI solutions}

\titlerunning{Towards Effective Immersive Technologies in Medicine}

\author{Aliaksandr Marozau\inst{1,2}\orcidID{0000-0003-3800-2899} \and
Barbara Karpowicz\inst{1,3}\orcidID{0000-0002-7478-7374} \and
Tomasz Kowalewski \inst{1,3}\orcidID{0009-0002-0542-9546}\and
Pavlo Zinevych \inst{1,3}\orcidID{0009-0008-9250-8712}\and
Wiktor Stawski \inst{1,3}\orcidID{0000−0001−8950−195X}\and
Adam Kuzdraliński\inst{1}\orcidID{0000-0003-2383-1950} \and
Wiesław Kopeć\inst{1,3}\orcidID{0000-0001-9132-4171}}
\authorrunning{A. Marozau et al.}

\institute{Polish-Japanese Academy of Information Technology, Warsaw, Republic of Poland \and
Proven Solution, Dubai, United Arab of Emirates
\email{morozovmdhealth@gmail.com} \and 
XR Space \url{https://xrspace.xrlab.pl/}
}

\maketitle             
\begin{abstract}
\begin{sloppypar}
Mixed Reality (MR) technologies such as Virtual and Augmented Reality (VR, AR) are well established in medical practice, enhancing diagnostics, treatment, and education. However, there are still some limitations and challenges that may be overcome thanks to the latest generations of equipment, software, and frameworks based on eXtended Reality (XR) by enabling immersive systems that support safer, more controlled environments for training and patient care.
Our review highlights recent VR and AR applications in key areas of medicine. In medical education, these technologies provide realistic clinical simulations, improving skills and knowledge retention. In surgery, immersive tools enhance procedural precision with detailed anatomical visualizations. VR-based rehabilitation has shown effectiveness in restoring motor functions and balance, particularly for neurological patients. In mental health, VR has been successful in treating conditions like PTSD and phobias.
Although VR and AR solutions are well established, there are still some important limitations, including high costs and limited tactile feedback, which may be overcome with implementing new technologies that may improve the effectiveness of immersive medical applications such as XR, psychophysiological feedback or integration of artificial intelligence (AI) for real-time data analysis and personalized healthcare and training. 

\end{sloppypar}

\end{abstract}

\section{Introduction}
Modern medicine is rapidly evolving with the introduction of immersive technologies such as virtual reality (VR), augmented reality (AR), and eXtended Reality (XR) that form together a mixed reality continuum (MR), which are transforming diagnostics, treatment, specialist training, and rehabilitation. These technologies provide advanced capabilities for creating immersive, interactive, three-dimensional environments, unlocking new possibilities for their application in medical practice \cite{senbekov2020recent}.

VR simulates virtual spaces that may resemble the real world, while AR overlays digital objects onto the physical environment \cite{desselle2020augmented}. XR combines elements of VR and AR, allowing real-time interaction between virtual objects and the real world. Therefore, XR provides a natural convergence of the former two well established technologies and can also be seen as a comprehensive functional equivalent to the Mixed Reality continuum.
\begin{sloppypar}
Immersive technologies are particularly valuable for enhancing patient-professional interactions and providing safe and replicable training and treatment conditions. They enable realistic simulations, making educational and therapeutic programs potentially more immersive, interactive, and effective \cite{kopec2023co,karpowicz2021intergenerational}. Additionally, VR and AR allow for innovative approaches in patient rehabilitation, such as restoring motor functions or providing psychological support, making these technologies powerful tools in various medical fields. However, effective and reliable training can benefit from real-time feedback, which is one of the foundations of our XR Framework. Based on our previous experience with experimental lab studies on multimodal data acquisition and processing in immersive systems, we are examining the potential of application of objective measurements and feedback in various application of well established VR/AR technologies towards more comprehensive XR and AI-based medical systems.
\end{sloppypar}
Therefore, having in mind constantly growing technologies related to immersive systems, in the following sections we discuss examples of successful and promising solutions applied in selected areas of medical application, including therapy, rehabilitation, surgery, and medical education, as discussed in the following sections.

\section{Immersive systems in medical student education}

Immersive technologies are transforming medical education by creating interactive and safe learning environments that simulate real clinical scenarios, allowing students and professionals to develop skills without risk to patients.

The American Heart Association (AHA) emphasizes the potential of using innovative educational strategies, such as immersive technologies, to improve resuscitation training. Methods such as gamification and virtual reality can complement existing training and enhance participants' performance, while also improving survival rates after cardiac arrest. These innovative strategies serve as an addition to traditional educational approaches, helping to improve learning and knowledge retention \cite{cheng2018resuscitation}.

Studies also demonstrate that VR and AR enhance spatial understanding of anatomical structures, offering particular benefits to students who might struggle with spatial skills. For example, VR can improve student engagement and understanding of complex anatomy, making it a valuable addition to educational programs, especially when access to physical specimens is limited \cite{moro2021virtual}.

VR technology allows learners to gain clinical experience safely, make mistakes, and learn from them while promoting autonomous learning and better material retention. Additionally, virtual simulations can reduce costs for educational institutions by minimizing the need for physical resources \cite{pottle2019virtual}.

While VR cannot replace all types of training, such as hands-on procedures requiring tactile feedback, it is effective for procedural algorithm learning. Furthermore, as character visualization quality and AI capabilities advance, VR may eventually be able to simulate complex communication and emotional skills \cite{pottle2019virtual}.

\section{Immersive systems in nursing education}

Current VR systems have proven to be effective in training nurses to interact with patients and perform various procedures. A meta-analysis demonstrated that VR training significantly enhances nursing knowledge and confidence in areas such as patient care for chronic conditions and specific procedures like catheterization and injections \cite{chen2020effectiveness}.

Studies also show that VR training contributes to a better understanding of clinical scenarios and increases engagement among nurses. Although some studies did not find substantial improvements in practical skills compared to traditional methods such us mannequins, VR still shows promise in boosting knowledge and self-assessment abilities \cite{kim2019effects}.

In a study with third-year nursing students, VR was reported to make learning more engaging and memorable compared to traditional methods. Students noted that VR helped maintain focus, provided access to a wider variety of clinical scenarios, and assisted with practical skills, from catheterization to suturing \cite{saab2021incorporating}. However, while VR is beneficial for training, it cannot replace the essential human interaction in nursing education.

\section{Immersive systems in surgery}

Immersive systems have proven valuable in surgery, a field that relies heavily on manual skills. VR simulators allow surgeons to practice complex procedures and emergency scenarios in a safe environment, reducing risks to patients and enhancing manual skills under high-stress conditions \cite{desselle2020augmented}.

For example, studies have explored the effectiveness of VR in training medical students and young surgeons in orthopedics, specifically in performing intramedullary nail (IMN) procedures on the tibia \cite{orland2020does,blumstein2020randomized}. In one study, 25 first- and second-year medical students were divided into three groups: one group used only technical guides, another group used VR, and the third group used both VR and technical guides. All participants performed the procedure on a synthetic bone model after 10-14 days of preparation \cite{orland2020does}.

As a result, a higher percentage of participants who used VR (with or without the guide) successfully completed the procedure compared to those who used only the guide (6 out of 8 and 7 out of 9, respectively, versus 2 out of 8). Participants in the VR groups made fewer errors compared to the group that used only the guide (an average of 3.2 errors in the VR groups versus 5.7 errors in the control group). Moreover, participants in the VR groups completed the procedure faster (on average, 19-18 minutes versus 24 minutes in the control group).

In another study, 20 students were divided into two groups: VR and standard guide. The participants performed the procedure on the SawBones simulator. The VR group completed the procedure 147 seconds faster than the guide group (615 seconds versus 762 seconds). Additionally, the VR group performed 38\% more correct steps than the control group (VR: 63\% versus 25\% correct steps). The VR group also received significantly higher scores in all categories (e.g., instrument handling, knowledge of the procedure) compared to the control group \cite{blumstein2020randomized}.

VR is particularly useful for training in laparoscopic and robot-assisted surgery. Research shows that VR simulators effectively transfer skills to real surgical settings, correlating strongly with performance in actual operations \cite{schmidt2021virtual}.

Beyond training, VR and AR are increasingly applied in surgical planning and intraoperative guidance. AR enables real-time interaction with 3D models of patient anatomy, improving spatial awareness and the precision of interventions. For example, using AR, surgeons can overlay 3D images of patient anatomy, like CT or MRI scans, onto the patient’s body during surgery, enhancing visualization and potentially reducing operation time \cite{desselle2020augmented}.

A notable example involves limb reconstruction surgeries using Microsoft HoloLens, which allows surgeons to view angiographic data overlaid onto the patient’s body. This tool enhances surgical accuracy while preserving sterility through hands-free controls. The device has shown promise in decreasing anesthesia time and improving surgical outcomes, as well as in training and providing remote support \cite{pratt2018through}.

AR-assisted surgeries further demonstrate enhanced precision and safety. For instance, in a shoulder replacement surgery, surgeons overlaid patient anatomy onto the body and received real-time feedback from remote colleagues, underscoring AR's potential in collaborative procedures \cite{gregory2018surgery}.

When preoperative 3D data are superimposed on the patient, surgical outcomes can surpass those using only 2D images. AR’s benefits extend to orthopedic, neurosurgery and hepatobiliary surgeries, where stable anatomical structures enable consistent overlay accuracy \cite{desselle2020augmented}.

Additionally, multiplayer applications in AR and VR now support collaborative review of imaging data, enabling remote teams to plan surgeries and visualize complex 3D models, which enhances diagnostic precision and procedural planning.

\section{Immersive systems in rehabilitation}

Immersive reality has seen widespread application in rehabilitation across various medical fields. A meta-analysis of 20 studies involving 518 patients with neurological, orthopedic, geriatric, or pediatric conditions showed that VR training — both specialized and game-based — had a positive impact on upper limb function in patients with neurological disorders. Specifically, VR therapy yielded better results on the Fugl-Meyer scale compared to traditional treatments, and significantly improved balance in patients with neurological conditions, as measured by the Berg Balance Scale and the Timed Up and Go test. However, no notable differences were found between VR and traditional methods regarding hand dexterity or gait improvements \cite{rutkowski2020use}.

A separate systematic review of eight randomized controlled trials with 805 knee arthroplasty patients found that VR rehabilitation significantly reduced pain in the first month post-surgery and showed improved WOMAC index scores for joint function. VR also enhanced HSS scores within the first two to three months, indicating accelerated recovery and pain reduction \cite{peng2022virtual}.

VR is also studied for rehabilitation in neurological disorders. A review of 14 trials with 524 Parkinson's patients demonstrated that VR improved balance significantly, as shown by the Berg Balance Scale and Activities-Specific Balance Confidence scale. Although VR led to slight gains in walking speed, no significant differences were noted in walking stability or quality of life compared to traditional rehabilitation methods \cite{kwon2023systematic}.

In stroke rehabilitation, VR has been shown to increase patient motivation and engagement, enhancing adherence to treatment and improving outcomes. VR’s immersive nature fosters a sense of presence, allowing patients to practice daily tasks in realistic scenarios, which increases functional independence. Traditional rehabilitation lacks this level of engagement and often results in reduced motivation. Studies also note that VR stimulates neuroplasticity—critical for motor function recovery in stroke patients \cite{aderinto2023exploring}.

A promising area is the use of VR and AR for training patients to use prosthetics that read nerve impulses. In a study with five participants, VR and AR were used to simulate tasks for upper limb prostheses. VR yielded higher efficiency and accuracy, while AR facilitated quicker task completion, showing that both technologies can complement each other in prosthetics training \cite{sun2021comparison}.

\section{Immersive systems in mental health}

Virtual reality has shown promising applications in mental health. Research indicates that virtual reality exposure therapy (VRET) is as effective as traditional exposure therapy for treating specific phobias, often preferred by patients due to controlled, repeatable environments that help reduce anxiety \cite{emmelkamp2021virtual}.

VR is also effective in treating post-traumatic stress disorder (PTSD), particularly among combat veterans. Patients undergoing virtual simulations of combat scenarios report reduced PTSD symptoms compared to control groups. A systematic review of 11 studies involving 438 patients demonstrated significant PTSD symptom reductions in VRET groups compared to waitlisted controls, with comparable results to traditional PTSD therapies. VRET is also deemed safe, with minimal risk of symptom exacerbation \cite{eshuis2021efficacy}.

Moreover, VR is used in the therapy of children with autism spectrum disorder (ASD) and attention deficit hyperactivity disorder (ADHD). Technology enables the creation of controlled environments, where children can practice skills such as attention control, speech development, and motor skills. VR training is actively used to enhance social and communication skills in children and adults with ASD. Studies have shown moderate success, especially in patients with high-functioning autism. For instance, young adults with high-functioning autism showed improvements in emotion recognition and theory of mind after VR training. Some studies describe changes in brain activity related to emotion processing and social situations, suggesting potential neuroplasticity induced by VR therapy \cite{emmelkamp2021virtual}.

\section{VR in pain management and stress reduction for patients}

The use of virtual reality for pain relief and stress reduction during medical procedures is becoming an increasingly popular method. The mechanisms by which VRET reduces pain can be explained by several factors. First, VR creates an immersive environment that helps divert the patient's attention away from painful sensations. This is achieved by immersing patients in virtual worlds, which shifts their focus from nociceptive stimuli (pain signals) to non-painful stimuli, thereby weakening the perception of pain. Second, VR alters emotional responses, reducing anxiety and discomfort associated with pain. This further decreases the subjective experience of pain, particularly in patients with acute pain symptoms. Third, the immersive environment created by VR stimulates non-painful signals and enhances the activity of neural pathways responsible for distraction from pain. Unlike medications, which affect the transmission of pain signals, VR works on a cognitive-emotional level, reducing the focus on pain and increasing tolerance \cite{huang2022using}.

VR is particularly effective for acute pain associated with burn care, needle procedures, and thermal pain exposure. For chronic pain conditions, such as lower back pain and cancer-related pain, no significant differences were found between VR therapy and standard treatments \cite{huang2022using}.

\section{Conclusions, challenges and future prospects}
Immersive systems based on VR and AR technologies are well established in medical education, training, and intervention (therapy). Despite the many advantages presented above, VR and AR still face some important challenges. Some of them can be solved by moving well established VR/AR applications towards XR-based solutions, which can provide better convergence between virtual and real elements of effective, comprehensive immersive environments, along with multimodal objective psychophysiological measurements and feedback in real-time combined with advanced AI-based data processing. One of the promising fields of improvement based on XR is related to limited tactile feedback in VR which is one of the main limitations in effective educational and training application. Another important element is related to personalization of education, training and therapy, which may benefit from real-time objective psychophysiological feedback as well as AI-based subsystems support in various aspects, from data processing to personalized content generation. In particular, recent advancements of our XR Framework, based on our previous experimental studies on effective multimodal psychophysiological data acquisitions in immersive systems, provide promising insights that medical applications may also benefit from real-time psychophysiological feedback towards better effectiveness of learning, training, and intervention processes \cite{kopec2023co}. On the other hand, some of the authors are also involved in the development of immersive systems that include adaptive therapeutic environments with virtual content creation, where AI generates real-time content tailored to the therapeutic needs of each patient. Such solutions like XR Framework or AI-based content generation systems may also reduce costs of customization of immersive environments for more effective, tailored and personalized medical application in education, training, and intervention.

\section*{Acknowledgments}
The authors did not receive any specific grant from funding agencies in the public, commercial, or not-for-profit sectors for this article.

The authors have no competing interests to declare that are relevant to the content of this article.

\bibliographystyle{bibliography/splncs04}
\bibliography{bibliography/bibliography}

\end{document}